%% file: NXBProbability.tex
\documentclass[fleqn,usenatbib,onecolumn]{mnras}

\usepackage[T1]{fontenc}

\DeclareRobustCommand{\VAN}[3]{#2}
\let\VANthebibliography\thebibliography
\def\thebibliography{\DeclareRobustCommand{\VAN}[3]{##3}\VANthebibliography}

\usepackage{graphicx}	
\usepackage{amsmath}	
\usepackage{amssymb}	
\usepackage{newtxtext,newtxmath}

\title[Probabilistic X-ray Background Removal]{A Probabilistic Method of Background Removal for High Energy Astrophysics Data}

\author[S.~Ehlert et al.]{
S.~Ehlert$^{1}$\thanks{E-mail: steven.r.ehlert@nasa.gov},
C.~T.~Chen$^{2}$,  
D.~Swartz$^{2}$,
R.~C.~Hickox$^{3}$, 
A.~Lutovinov$^{4}$,
A.~Semena$^{4}$,
\newauthor
R.~Krivonos$^{4}$, 
A.~Shtykovsky$^{4}$, \&
A.~Tkachenko$^{4}$
\\
$^{1}$ NASA Marshall Space Flight Center, Huntsville, AL 35812, United States\\
$^{2}$Universities Space Research Association, Huntsville AL 35805 USA\\
$^{3}$Department of Physics and Astronomy, Dartmouth College, 6127 Wilder Laboratory, Hanover NH 03755, USA\\
$^{4}$Space Research Institute, 84/32 Profsouznaya str., Moscow 117997, Russian Federation\
}

\date{Accepted XXX. Received YYY; in original form ZZZ}

\pubyear{2021}

\newcommand{\lae}{\mathrel{<\kern-1.0em\lower0.9ex\hbox{$\sim$}}}
\newcommand{\gae}{\mathrel{>\kern-1.0em\lower0.9ex\hbox{$\sim$}}}

\input{units}

\begin{document}
\label{firstpage}
\pagerange{\pageref{firstpage}--\pageref{lastpage}}
\maketitle





\begin {abstract}
We present a new statistical method for constructing background subtracted measurements from event list data gathered by X-ray and gamma ray observatories. This method was initially developed specifically to construct images that account for the high background fraction and low overall count rates observed in survey data from the Mikhail Pavlinsky ART-XC telescope aboard the \textit{Spektrum R\"{o}ntgen Gamma (SRG) } mission, although the mathematical underpinnings are valid for data taken with other imaging missions and analysis applications. This method fully accounts for the expected Poisson fluctuations in both the sky photon and non X-ray background count rates in a manner that does not result in unphysical negative counts. We derive the formulae for arbitrary confidence intervals for the source counts and show that our new measurement converges exactly to the standard background subtraction calculation in the high signal limit.  Utilizing these results, we discuss several variants of images designed to optimize different science goals for both pointed and slewing telescopes. Using realistic simulated data of a galaxy cluster as observed by ART-XC we show that our method provides a more significant and robust detection of the cluster emission as compared to a standard background subtraction. We also demonstrate its advantages using real observations of a point source from the ART-XC telescope. These calculations may have widespread applications for a number of source classes observed with high energy telescopes.

\end {abstract}


\section{Introduction} \label{s:intro}
The presence of background events is  ubiquitous in X-ray and gamma ray observations of the sky, and accounting for the presence of background is usually essential for measuring the properties of sources accurately. These telescopes, however, often reside in a unique region of parameter space. Because of the expected X-ray and gamma ray photon flux rates, there may only be a handful of observed events in a region of interest. Even after a relatively long exposure, many detector pixels may ultimately not register any events. These realities lead to a highly discretized image where the counts per pixel is $\sim \mathcal{O}(1-10)$. This regime can be challenging for analysis, especially since these small total counts include two distinct event source populations: the incident astrophysical photons and non astrophysical background events. For the purposes of this paper, we define the non X-ray background (NXB) as the collection of events originating from sources  other than photons emitted by astrophysical sources. This includes (but is not limited to) charged particles impacting the detectors from any direction, secondary photons produced by charged particle interactions with the telescope hardware, and electronic readout noise. These events are frequently difficult to distinguish from ``real'' astrophysical photons on an event-by-event basis, but typically have different spectral and spatial distributions in the detector. Separating the astrophysical photons from the background events is of crucial importance to properly interpreting the data, detecting sources, or estimating the emission of diffuse and unresolved sources.

How background is accounted for in an analysis of high energy telescope data strongly depends on the science case and nature of the observational data. The background counts can in many cases be the dominant limitation and  source of systematic uncertainty and errors for studies of faint, diffuse sources such as the the hot gas at the outskirts of galaxy clusters \citep[e.g.][and references therein]{Reiprich2013} or the cosmic X-ray background \citep{Hickox2006}.

The simplest method for dealing with background is subtraction - otherwise identical science and background products (e.g. images or spectra) are constructed and the net signal is created as the difference of the two. This method has a few serious drawbacks for data in the X-ray and gamma ray regimes. The first and in some ways the most obvious is that the expected background counts can often exceed what is observed due to routine Poisson fluctuations. In this case the net counts are in fact negative, which is unphysical. These negative pixels/channels can be ignored or modified to be non-negative, albeit at the cost of losing information about the source. A related concern is that the large Poisson fluctuations expected for observations with only a few counts can lead to regions where noise can be misinterpreted as signal. Both of these concerns can be further compounded by the fact that high energy telescopes generally do not point at a single sky position during an observation the telescope is usually dithering around or slewing across a particular sky position during the nominal exposure, complicating the mapping between detector coordinates (where many of the NXB events are often generated) and sky coordinates (where the background of interest usually needs to be measured). One final challenge with a simple subtraction from a statistics perspective is that while the sum of two Poisson distributed random variables is another Poisson distribution, the difference is not\footnote{The difference of two Poisson distributed random variables in fact follows a Skellam distribution \citep{Skellam1946}.}.

More sophisticated and rigorous methods of accounting for the background around sources often require the use of a model for the source in question. For example, the survey for sources in the \textit{Chandra} Deep Field South \citep[e.g.][ and references therein]{Luo2017} compares counts within a model of the Point Spread Function (PSF) to those within a larger source-free region to calculate the 
significance of emission at the positions of candidate sources. Another method for identifying sources includes involving images generated from event lists with a filter whose properties involve assumptions of the telescope PSF and background, such as the analysis described in \cite{Pavlinsky2021Cat}. Such methods are less frequently utilized for extended and diffuse sources, as implementing the source models for extended sources in such methods is more uncertain and cumbersome.  


The data from Mikhail Pavlinsky ART-XC (hereafter ART-XC) telescope on board the \textit{Spectrum-R\"{o}ntgen-Gamma (SRG)} mission also operate in a background limited regime. The seven telescopes comprising this instrument are currently performing the first high spatial resolution ($\sim 50^{\prime\prime}$) all-sky survey of the X-ray sky in the $4- 30 \keV$ bandpass. The rapid motion of ART-XC across the sky results in ART-XC's event lists having only short exposure times at each sky position, with only a handful of photons expected to arrive with each pass. Compared to the other X-ray astronomy missions described above, background events in the ART-XC detector are a significantly higher fraction of the total events that cannot be easily ignored or neglected. Removing background counts from these data is clearly subject to the limitations mentioned in the above paragraphs.

In order to address these challenges and produce additional meaningful measurements of the sky with the ART-XC event list data, we present a new statistical method for removing background events from X-ray images and estimating the sky photon count rates in detector coordinates. Our new method is very similar to a simple background subtraction but is not subject to the most serious drawbacks and limitations of that method. It is therefore especially well suited for analyzing diffuse and unresolved sources at low signal-to-noise in a model-independent fashion.  The fundamental assumption behind this method is that, given the data available for each event, it is impossible to definitively identify any individual event as an background event or a photon from an astrophysical source (hereafter a sky photon). We can therefore only assign a probability as to it originating from the background or the sky. Although this methodology was developed specifically with ART-XC event lists (where the total counts per detector pixel are generally $\sim 30$ or fewer and dominated by background events) in mind, the mathematical underpinnings of this method are valid for nearly all high energy astrophysics missions that observe data as discrete photon events. This methodology may have many useful potential applications for other X-ray and gamma ray telescope data currently limited by low signal and high background count rates. 

This paper is organized as follows: Section 2 discusses the mathematical foundations of how to estimate the sky photon counts in a given detector region, and Section 3 describes where and when these methods provide advantages over past practices. Section 4 describes some tests using Monte Carlo simulations to show that our new estimator is unbiased and compare its performance to a standard background subtraction calculation.  Section 5 provides more expansive details as to how these calculations can be used to produce images for both pointed and survey observations, and Section 6 shows images produced using these new calculations on both real and simulated ART-XC data. Section 7 provides concluding remarks.

\section{Estimating the Sky Photon Rate} \label{s:skymath}

Our process hinges on the following assumptions about the X-ray observation event list, and unless otherwise noted all variables are considered for an individual region on the detector or sky as small as an individual detector pixel:

\begin{enumerate}

\item The event list has already undergone a full battery of cleaning and processing steps. These steps may include restrictions on the energy band, event times, data quality flags, or region of the detector. While this is not strictly necessary from the standpoint of the mathematics we are about to describe, this method is not designed to further refine the parameter space of event properties.
\item The background events and sky photons in the event list are drawn from independent Poisson processes with expectation values of $\mu_{N}$ and $\mu_{S}$, respectively. These two variables are in units of integrated counts in that detector region over the integrated exposure. While the calculations described below do not make any assumptions about the relative values of these two expectation values, in practice we are most concerned with the regime where $\mu_{N} > \mu_{S}$ and $\mu_{N} \gtrsim 0.1$. 
\item  An estimate for $\mu_{N}$ for the event list and detector region in question can be derived from either these same data or a different data set \textit{a priori}. Precisely how this calculation is made, what its value is, and which data it uses will be strongly dependent on the telescope in question. As a concrete example, the expected background counts for a given science observation might be estimated using ``blank-sky'' observations subject to the same quality cuts.  
\item For the region in question, we observe a total of $k$ counts, of which $k_{N}$ originate from the background and $k_{S}$ are sky photons. The exact values of $k_{N}$ and $k_{S}$ cannot be directly measured, but they are obviously both non-negative integers that sum to $k$. Our final estimate will need to account for all possible states of the observed counts into background and sky photon components (i.e. $k_{N}$ taking all values from $0$ to $k$). 
\end{enumerate}

The value of scientific interest to measure is $\mu_{S}$, which is the expectation value for the number of events in this particular region that originate from sky photons. Using the variables, notation, and assumptions described above, a simple background subtraction calculation would estimate $\mu_{S}$ as 

\begin{equation}\label{eq::BasicSubtract}
    \mu_{S} = k - \mu_{N}
\end{equation}

When considering the regime where $\mu_{N} \sim k$ and $k \sim \mathcal{O}(10)$, the result of Equation \ref{eq::BasicSubtract} has a high probability of resulting in a negative value for $\mu_{S}$ due to the expected Poisson fluctuations of the background (i.e there is a high probability of observing fewer counts than the expectation value of the background). Furthermore, the Poisson fluctuations of both variables can greatly skew this estimate of the expected sky counts when $\mu_{S} \sim \mu_{N}$. These are the regimes for which this calculation are designed to address.  Our treatment of our two Poisson processes will be Bayesian - rather than assume that the observed number counts $k$ is a randomly distributed variable about ``known'' values of $\mu_{S}$ and $\mu_{N}$, we will instead keep the expected background counts $\mu_{N}$ fixed while calculating the probabilities of $\mu_{S}$ given that $k$ counts were observed. 

The fundamental formula at the heart of this entire calculation is the canonical Poisson probability mass distribution. For a known value of $\mu$, we denote the probability of observing $j$ counts as 

\begin{equation}\label{eq::SinglePoisson}
P(j \vert \mu ) = \frac{ e^{-\mu} \mu^{j}}{j !}
 \end{equation}

\section{The Sum of Two Poisson Distributions - The Bayesian Perspective}

Because of the assumptions we are making and the relative mathematical simplicity of the Poisson distribution, we can use Bayesian statistics to construct a posterior probability distribution calculation that assumes the broadest and flattest possible prior\footnote{In other words, the posterior is directly proportional to the likelihood rather than the product of prior and likelihood.} on our only ``unknown'' parameter $\mu_{S}$. Assuming that the prior distribution of $\mu_{S}$ is a constant across the entire feasible domain, i.e. $\mu_{S} = a \thinspace \forall \thinspace \mu_{S} \in \left( 0, \infty\right)$, may at first glance seem excessively broad. As we will now demonstrate, however,
an exact analytic formula for an arbitrary central posterior density interval can be calculated under these assumptions.

Since the sum of two Poisson distributed variables is itself a Poisson distributed variable, we can begin by defining the likelihood function for this model $\mathcal{L}( \mu_{S} \vert k, \mu_{N}) d \mu_{S}$ as follows

\begin{equation}\label{eq::DoublePoissonToRenorm}
     \mathcal{L}( \mu_{S} \vert k, \mu_{N}) d\mu_{S}= \frac{e^{-\left(\mu_{N} + \mu_{S}\right)} \left(\mu_{S} + \mu_{N}\right)^{k} }{k!} d\mu_{S}
\end{equation}

\noindent which is effectively identical to the probability mass function in Equation \ref{eq::SinglePoisson}. By requiring the integral of this function over $\mu_{S}$ to be equal to unity, we determine the posterior PDF as

\begin{equation}\label{eq::DoublePoissonRenormed}
     \mathcal{P}( \mu_{S} \vert k, \mu_{N} ) d\mu_{S}= \frac{e^{-\left(\mu_{N} + \mu_{S}\right)} \left(\mu_{S} + \mu_{N}\right)^{k} }{\Gamma_{U}(k +1, \mu_{N})} d\mu_{S}
\end{equation}
\noindent where $\Gamma_{U}(k +1, \mu_{N})$ is the incomplete upper gamma function, replacing the $k!$ term in the denominator in order to normalize the integral over $\mu_{S}$ instead of the sum over $k$. A simple derivative calculation shows that this posterior PDF is maximized at either $\mu_{S} = k - \mu_{N}$ whenever this difference is non-negative or $\mu_{S} =0$ whenever $k - \mu_{N} < 0$. We also note that, in the absence of any background (i.e. $\mu_{N} = 0$), that the denominator would be exactly equal to $k!$, leaving Equation \ref{eq::DoublePoissonToRenorm} unchanged.  The expectation value of $\mu_{S}$ for this posterior PDF is given as 

\begin{equation}\label{eq::MeanValueIntegral}
\left< \mu_{S} \right> = \int_{0}^{\infty} \mu_{S} \mathcal{P}( \mu_{S} \vert k, \mu_{N}) d\mu_{S}
\end{equation}
\noindent and can also be calculated exactly as 
\begin{equation}\label{eq::MeanValue}
 \left< \mu_{S} \right> = \frac{\Gamma_{U}(k+2,\mu_{N}) - \mu_{N} \Gamma_{U}(k+1,\mu_{N}) }{\Gamma_{U}(k+1,\mu_{N})}
\end{equation}

\noindent We can also calculate the cumulative distribution function $CDF(x)$ as 
 
\begin{equation}\label{eq::CDF}
CDF(x) = \int_{0}^{x} \mathcal{P}( \mu_{S} \vert k, \mu_{N}) d\mu_{S}   
\end{equation}
\noindent The CDF has an exact analytic form, and can be written as

\begin{equation}\label{eq::Analytic}
\int_{0}^{x} \mathcal{P}( \mu_{S} \vert k, \mu_{N}) d\mu_{S} = \frac{\gamma_{L}(k+1,x + \mu_{N} ) - \gamma_{L}(k+1, \mu_{N} )}{\Gamma_{U}(k+1,\mu_{N})}   
\end{equation}
\noindent \noindent where $\gamma_{L}$ is the incomplete lower gamma function. Using this form, the median value of this posterior PDF can be determined by solving the equation $CDF(x) = \frac{1}{2}$ numerically. Similarly, this analytic formula also enables a central posterior density interval to be determined. The upper and lower bounds of an interval that contains a fraction $c \in (0,1) $ of the posterior PDF can be determined by solving $CDF(x) = \frac{1 \pm c}{2}$. As a concrete example, the lower and upper bounds of the $68\% \thinspace (1-\sigma)$ posterior density interval of $\mu_{S}$ can be calculated by setting the right hand side of Equation \ref{eq::Analytic} equal to 0.16 and 0.84, respectively.

\section{A Summary Statistic for the Posterior Distribution}

The results of the previous section suggest a few summary statistics that might serve as well-motivated point estimates for posterior PDF given $\mu_{N}$ and $k $. However, we have identified another summary statistic of of $\mu_{S}$ (which we denote as $\mu_{S}^{\star}$) that has better asymptotic behavior and lower bias than any of the estimators derived from the posterior PDF discussed thus far. In this section we will construct this point estimator and consider it in the context of the posterior PDF results of the previous section. 

To construct $\mu_{S}^{\star}$, we first consider an individual micro-state of the data where $k_{N}$ events originate from the background and $k_{S}$ events originate from they sky. For this micro-state, we can calculate the posterior PDF for $\mu_{S}$ given $k_{S}$ observed counts and a uniform prior over the domain of $\mu_{S} \in (0, \infty)$, $\mathcal{P}^{\prime} ( \mu_{S} \vert k_{S)}$ as the canonical Poisson function 

\begin{equation}
    \mathcal{P}^{\prime} ( \mu_{S} \vert k_{S}) d\mu_{S} = \frac{e^{-\mu_{S}} \mu_{S}^{k_{S}}}{k_{S}!} d\mu_{S}
\end{equation}
\noindent where we are able to simplify significantly as compared to Equation \ref{eq::DoublePoissonRenormed}. For this particular micro-state, we can determine a single-point estimate of $\mu_{S}$ (denoted as $\bar{\mu}_{S}(k_{N})$) by calculating the value of $\mu_{S}$ for which this posterior PDF is maximized. Solving the elementary calculation equation 
 
 \begin{equation}
 \frac{\partial \mathcal{P}^{\prime}( k_{S} \vert \mu_{S}) }{\partial \mathcal{\mu_{S} } } = 0
 \end{equation}
 
\noindent results in a simple and intuitive relationship between $\bar{\mu}_{S}$ and $k_{S}$
 \begin{equation}
     \bar{\mu}_{S}(k_{N}) = k_{S}  = k - k_{N} 
 \end{equation}

\noindent that is trivial to calculate for all values of $k_{S}$. We can also calculate the probability of observing $k_{N}$ background events given the expected background counts $\mu_{N}$
 
 \begin{equation}\label{eq::DoublePoisson}
  w(k_{N} \vert \mu_{N}, k ) =  P(k_{N} \vert \mu_{N}) = \frac{e^{-\mu_{N}} \mu_{N}^{k_{N}}}{k_{N}!}   
 \end{equation}
 
\begin{figure*}
     \centering
       \includegraphics[width=0.45\textwidth]{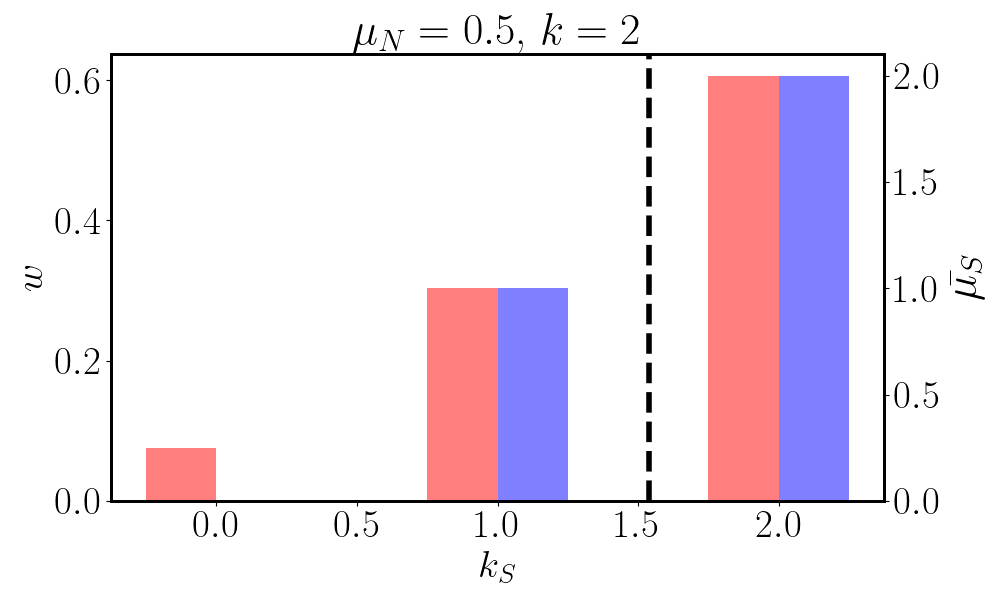}
    \includegraphics[width=0.45\textwidth]{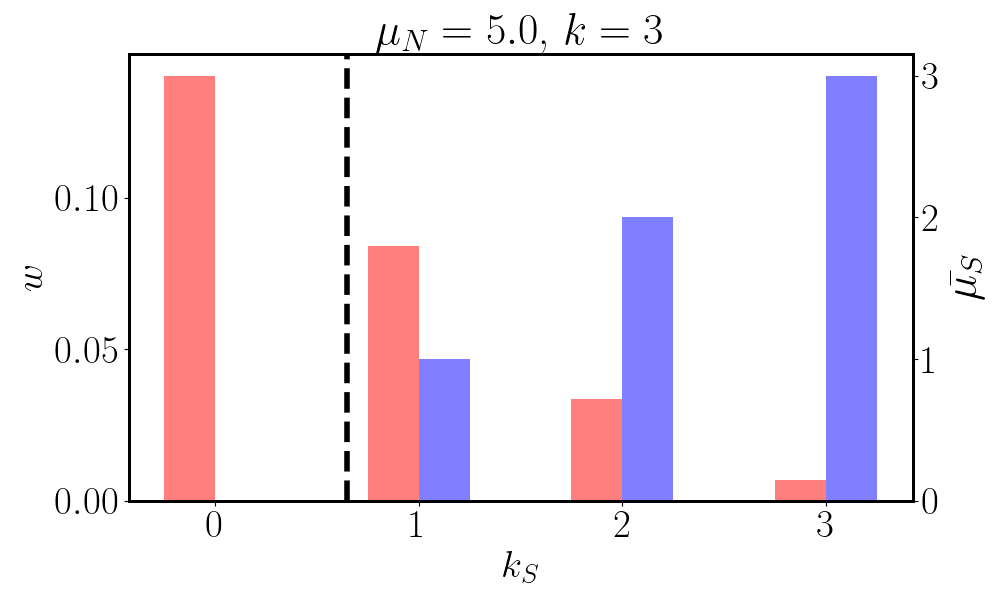}
    \includegraphics[width=0.45\textwidth]{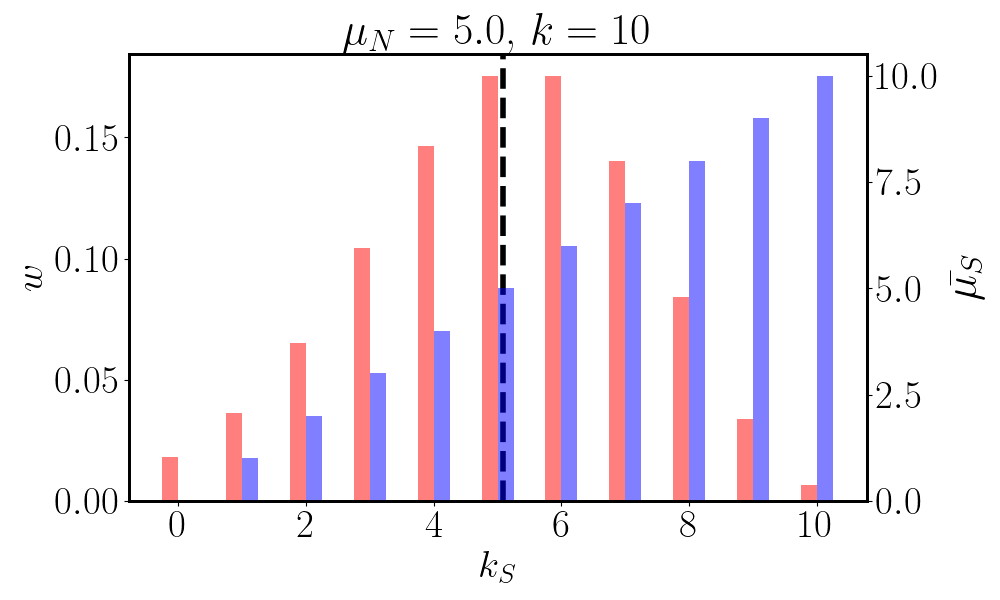}

    \caption{ Values of $\bar{\mu}_{S}$ (in blue, y-axis on the right side of the figure) and $w(k_{N} \vert \mu_{N}, \bar{\mu}_{S}, k )$ (in red, y-axis on the left side of the figure) as a function of the assumed number of observed sky counts ($k_{S}$ for each state). Three different values of the expected background and observed counts shown: The top left panel shows $\mu_{N} =0.5 $ and $k=2$. The top right panel shows $\mu_{N} = 5.0$ and $k=3$, a situation where simple subtraction would result in negative counts. The bottom panel shows $\mu_{N}=5.0$ and $k=10$. In all three figures the vertical dashed line denotes our new calculation for the expected sky counts, $\mu_{S}^{\star}$.   For all three cases it should be clear that many different configurations of the data have non-negligible probabilities, especially for the case where $\mu_{N} = 5.0$ and $k=10$ total counts.  }
    \label{fig::MuSBarWeights}
 \end{figure*}

 \noindent  Some example values of $\bar{\mu}_{S}$ and $w(k_{N} \vert \mu_{N}, \bar{\mu}_{S}, k )$ are shown in Figure \ref{fig::MuSBarWeights}. Our final measurement of the expected sky photon counts in this region, $\mu_{S}^{\star}$, is then calculated as a weighted average of the values of $\bar{\mu}_{S}$ over all possible values of $k_{N}$, where the weights themselves are the probabilities of observing the background counts in that micro-state
 
\begin{equation}
\label{eq::mustar}
\mu_{S}^{\star} = \frac{1}{V_{1}} \sum_{k_{N} = 0}^{k}  \bar{\mu}_{S}(k_{N}) \times w(k_{N} \vert \mu_{N}, k ) 
\end{equation}

\noindent with the variable $V_{1}$ being the sum of all of the weights

\begin{equation}
V_{1} =  \sum_{k_{N} = 0}^{k} w(k_{N} \vert \mu_{N}, \bar{\mu}_{S}, k )
\end{equation}

\noindent This weighted average takes into account all possible realizations as to how the $k$ observed counts in the region can be distributed into sky photons and background counts. It is also by construction positive definite irrespective of the values of $\mu_{N}$ and $k$. As we show in Appendix \ref{app::converge}, this probabilistic summary statistic also converges exactly to simple background subtraction in the regime where $k \gg \mu_{N}$. We finally use $\mu_{S}^{\star}$ to calculate the probability that each individual event originates with a sky photon, $p_{S}$, as 

\begin{equation}
p_{S}  = \frac{\mu_{S}^{\star}}{\mu_{N} + \mu_{S}^{\star}}
\end{equation}

\noindent Similarly, the probability for a single event being associated with an background event, $p_{N}$, is estimated as 

\begin{equation}
p_{N}  = \frac{\mu_{N}}{\mu_{N} + \mu_{S}^{\star}}
\end{equation}
\noindent We emphasize that these two probabilities clearly sum to unity. For data at very high signal-to-noise where $k \gg \mu_{N}$, one can simply utilize the result of simple background subtraction ($k - \mu_{N}$) in the place of $\mu_{S}^{\star}$, which can help remove the background events from images and other data products (see \S \ref{S::imagemake} for more details).   

\begin{figure*}
     \centering
    \includegraphics[width=0.45\textwidth]{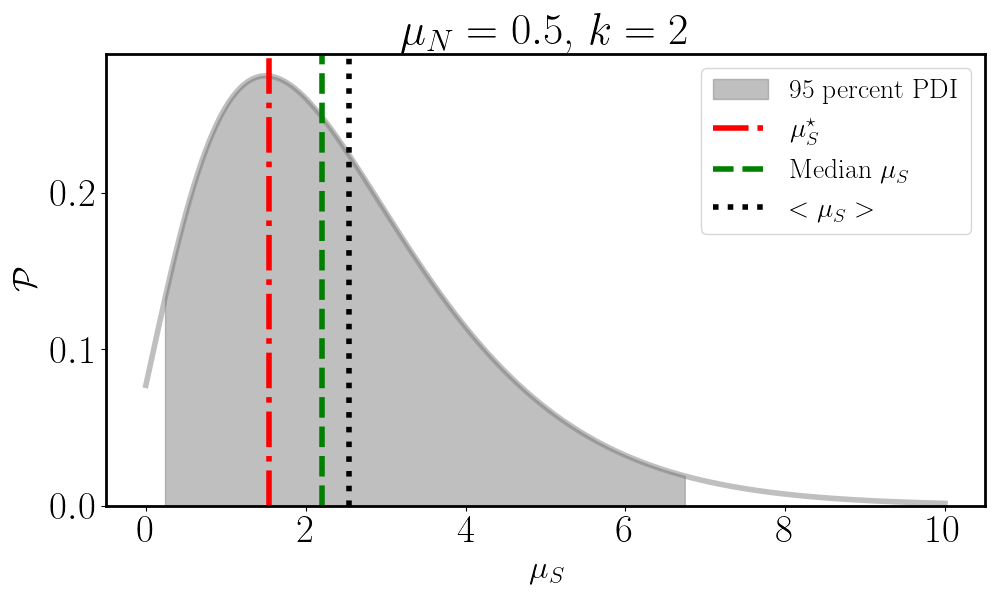}
    \includegraphics[width=0.45\textwidth]{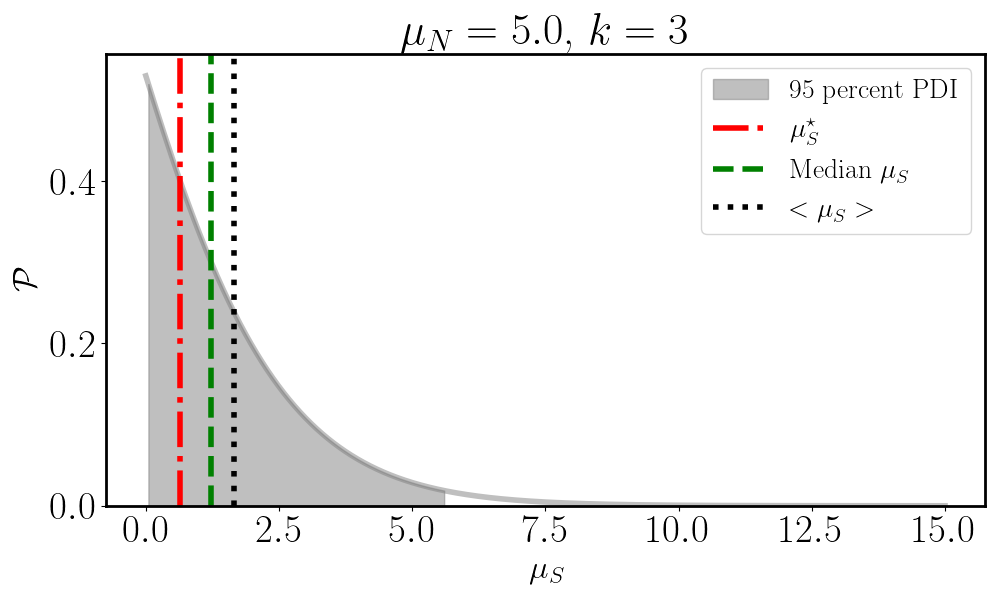}
    \includegraphics[width=0.45\textwidth]{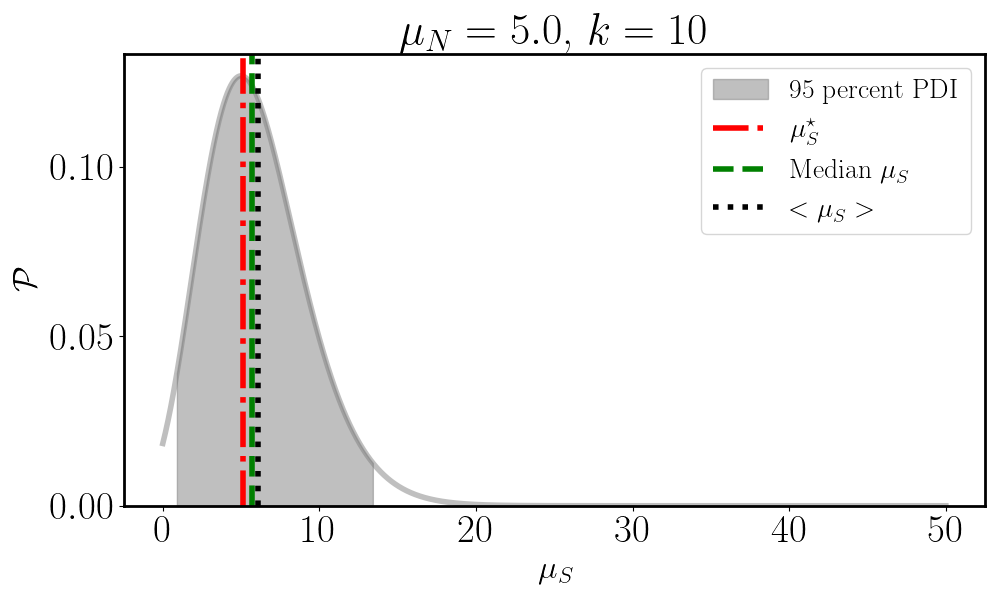}
    
    \caption{ Posterior probability density functions and $95\%$ posterior density intervals for three different configurations of $\mu_N$ and $k$. For all three figures, the region containing $95\%$ of the integrated probability is shaded in gray, while the derived value of $\mu_{S}^{\star}$ is denoted with a vertical dot-dashed red line. The median values are shown as dashed green lines, and the mean values (see Equation \ref{eq::MeanValue} ) are shown as the dotted black lines.   }
    \label{fig::ConfidenceIntervals}
 \end{figure*}

Three examples of the posterior PDF $\mathcal{P}$ and the corresponding $95\%$ posterior density intervals ($c=0.95$) for different values of $\mu_{N}$ and $k$ are shown in Figure \ref{fig::ConfidenceIntervals}. These figures show a few properties of these confidence intervals that hold generally, mainly that the intervals are highly asymmetric around our derived value of $\mu_{S}^{\star}$. This asymmetry arises immediately from the asymmetry in the underlying Poisson-like distribution. Our estimator is also systematically lower than the mean and median estimators of $\mu_{S}$. 

\begin{figure*} 
    \centering
    \includegraphics[width=0.75\textwidth]{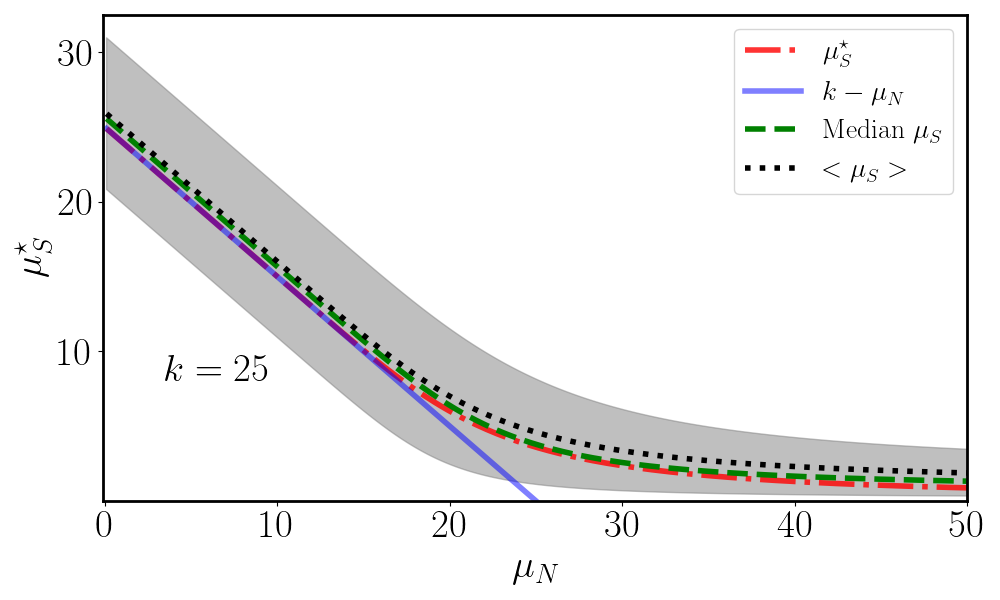}
    \caption{ The estimate sky event counts  $\mu_{S}^{\star}$ as a function of the expected number of background events $\mu_{N}$ for the case where $k=25$, in red. The shaded red region corresponds to the $68.3\%$ confidence interval around $\mu_{S}^{\star}$, as calculated using the results of Equation \ref{eq::Analytic}. The blue curve denotes the expectation from simple subtraction $\mu_{S}^{\star} = k -\mu_{N}$, which obviously results in unphysical negative counts when $\mu_{N} > k$.  The green dashed and black dotted lines correspond to the median (see Equation \ref{eq::Analytic} and mean (see Equation \ref{eq::MeanValue} ) values of $\mu_{S}$, respectively.  The estimate of $\mu_{S}^{\star}$ is by construction positive definite and converges to zero as the expected background gets much larger than the observed counts. The median and mean values show similar behavior but tend to be higher than $\mu_{S}^{\star}$ and $ k - \mu_{N}$ by approximately 1 count.  }
    \label{fig::MuSVMuN}
\end{figure*}

A plot showing how the value of our three summary statistics and the $68\%$ posterior density interval depends on $\mu_{N}$ for a total of $k=25$ events is shown in Figure \ref{fig::MuSVMuN}. One particularly convenient property of our new estimator $\mu_{S}^{\star}$ is that it converges exactly to simple subtraction when the observed counts are much higher than the background (see the Appendix for a proof of this), while the median and mean values are systematically higher. As the expected background increases (while keeping the total number of observed counts unchanged) our estimator $\mu_{S}^{\star}$ smoothly converges to zero as expected, although it remains positive for all values of $\mu_{N}$. This effect is due to the fact that the total observed counts become increasingly more likely to be a fluctuation in the background than evidence of a second independent source of events.   

\subsection{Regimes of Applicability}

While mathematically valid for any number of counts drawn from a Poisson distribution, this method should not completely supersede simpler background subtraction methods. When there are sufficient sky photon events, simple background subtraction and the Gaussian approximations are sufficiently accurate. At the same time, the probabilistic treatment can be  more time consuming to calculate at higher signal-to-noise. Unlike a simple subtraction calculation, a sum over all integers up to the total counts in that region is required to estimate the sky photon counts using this probabilistic method.  In other words, a region with $k$ events requires at least a factor of $k$ more arithmetic operations than standard background subtraction. For event lists that have $k \sim \mathcal{O}(100)$ counts per region this can become significantly slower than a simpler, more traditional background subtraction calculation with negligible gains in accuracy. 

Considering the numerical stability and accuracy of the calculation, the sum over all states is straightforward to calculate accurately for all values of $\mu_{N}$ and $k$. When the expected number of background counts greatly exceeds the total number of observed counts ($\mu_{N} \gtrsim 2k$), however, the ability to find roots for the CDF as calculated in Equation \ref{eq::Analytic} can be limited by the numerical precision to which the incomplete gamma functions can be computed. It may not be possible to numerically determine a posterior density interval in these cases. This is also the situation where the observations are a low-probability realization of the expected background in an absolute sense, so we encourage any users who find themselves in this situation to review their background models carefully.  

One final detail that was implicitly assumed for the calculations in \S \ref{s:skymath} is that the expected background counts in the region in question ($\mu_{N}$) is a value of order unity or larger. For situations where the expected background counts in the test region are significantly smaller than unity, the probability of any individual event being associated with the background becomes negligible and practically every event is treated as being from sky photons. Given the realities of total count rates for X-ray and gamma ray telescopes, those who wish to apply these calculations to real data should bin the event lists either in regions of detector space or time intervals in order to have an expected background counts of $\mathcal{O}(1)$, or at least to where there is a non-negligible probability of detecting at least one background event. Using nominal ART-XC values as a quantitative example as to why such binning is necessary, we will assume a background count rate of $\sim 10^{-4} \thinspace \mathrm{cts} \s^{-1} \thinspace \mathrm{pix}^{-1}$ for a detector with $1800$ pixels. If the data is broken up into $1 \s$ time intervals at the native detector resolution (i.e. with $\mu_{N} = 10^{-4}$), we would infer that practically every event is associated with a sky photon (for $k=1$ observed counts in a pixel, $\mu_{S}^{\star} = 0.9999$). The need for such binning is not unique to our new calculation -  simple background subtraction results in an identical value. In fact, given the design of this calculation bins an order of magnitude smaller than what may be needed for Gaussian approximations to hold can be used. It is beyond the scope of this paper to discuss the best practices for binning up the data for an general observatory, however, given that these choices will strongly depend both on the mission-specific details of the telescope as well as the science questions of interest.

\section{Performance of our new estimator}

\subsection{Bias}

One of the desirable properties of a simple subtraction calculation is that it is an unbiased statistical estimator of $\mu_{S}$ (see the Appendix for a simple proof of this claim), at least for situations where the difference is not restricted to non-negative values. Imposing the physical restriction that our estimator of $\mu_{S}$ must be non-negative will inevitably introduce some bias into our estimation. In order to investigate the bias for our new summary statistic $\mu_{S}^{\star}$, we utilize Monte Carlo simulations to estimate the expected bias compared to $k-\mu_{N}$. We also show, as a comparison, the equivalent bias calculations for the other two summary statistics presented earlier in this paper.   

We have run 25,000 Monte Carlo simulations of Poisson processes with randomly determined ground-truth values of $\mu_{S,true}$ and $\mu_{N,true}$ between 0.05 and 50. This range of parameter values is based on the arguments made in the previous section, since this calculation does not offer sizable advantages in either accuracy or computation speed compared to the simple subtraction method when the total counts exceed $\sim 100$.  For each realization of these two parameter values, we generate observed counts using Poisson distributions and calculate all four summary statistics. We calculate the average bias with respect to the true simulation input value of $\mu_{S}$.  

\begin{figure*}
    \centering
    \includegraphics[width=0.75\textwidth]{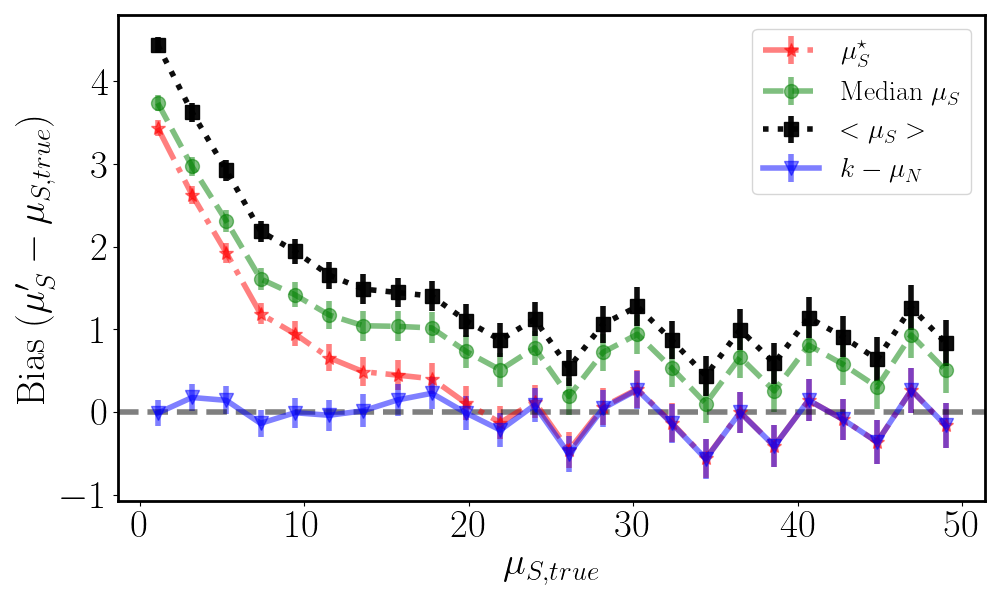}
     \caption{Comparisons of the bias between probabilistic and simple background subtraction calculations. These curves show the average bias as a function of the true input values of $\mu_{S}$. In addition to our new estimator $\mu_{S}^{\star}$ we also show the bias for the median $\mu_{S}$ and $<\mu_{S} >$ summary statistics derived in this work. While all three of the new summary statistics are biased compared to simple subtraction, our new estimator $\mu_{S}^{\star}$ is the least biased at all values of $\mu_{S}$ and converges to the simple subtraction result at high values of $\mu_{S}$.     }
    \label{fig::MethodComp}
\end{figure*}

The results of this exercise can be found in Figure \ref{fig::MethodComp}. As expected, all three of our new summary statistics are found to be biased, especially at the lowest values of $mu_{S}$. Of the three, only our new estimator $\mu_{S}^{\star}$ converges to zero bias at high values of $\mu_{S}$. This numerical convergence in the average bias is unsurprising given the mathematical convergence between the two estimators in this regime.  The median value of $\mu_{S}$ and mean value $<\mu_{S}>$ both converge to higher, non-zero values of bias around $\sim 0.6-1.2$ counts for all values of $\mu_{S}$ investigated here.  

\subsection{Central Tendency}

\begin{figure*}
    \centering
    \includegraphics[width=0.75\textwidth]{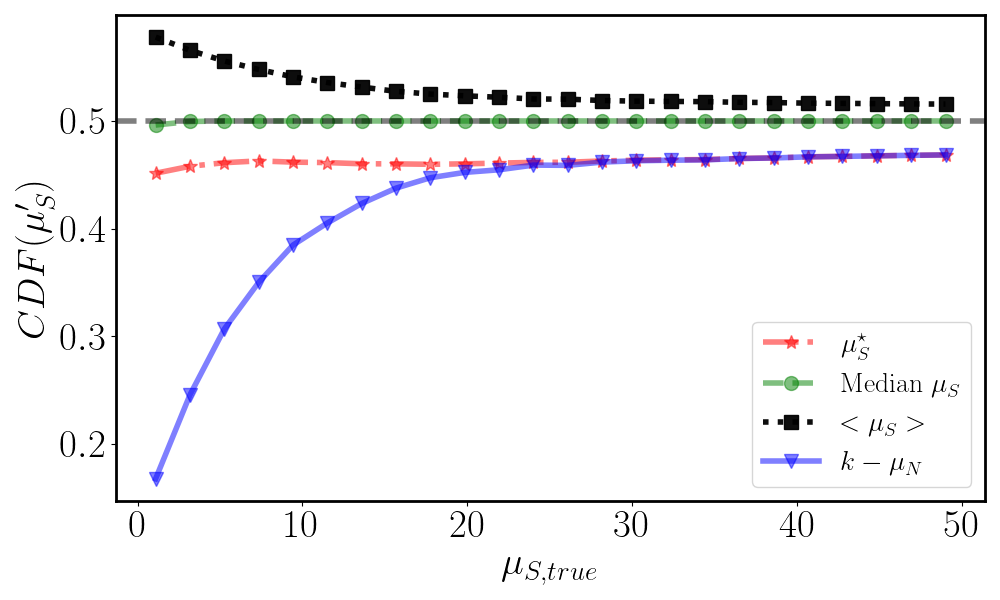}
     \caption{Comparisons of the CDF between probabilistic and simple background subtraction calculations. These curves show the average value of the CDF for each estimator as a function of the true input values of $\mu_{S}$. In addition to our new estimator $\mu_{S}^{\star}$ we also show the CDF for the median $\mu_{S}$ and $<\mu_{S} >$ summary statistics derived in this work.     }
    \label{fig::MethodCompCDF}
\end{figure*}

The performance of our new estimator cannot be solely judged on the basis of its bias. Another dimension for which a particular summary statistic can be tested is whether or not its value is representative of the underlying distribution. To investigate this, we calculated the average value of the posterior CDF (see Equation \ref{eq::CDF}) at each of the four estimators as a function of the ground-truth value of $\mu_{S}$ used as simulation input. By construction our median $\mu_{S}$ estimator will have an average CDF value of 0.5.  The results, shown in Figure \ref{fig::MethodCompCDF}, demonstrate that our new estimator has a nearly constant average CDF value of $\sim 0.46$. The average CDF for the simple subtraction calculation\footnote{When simple subtraction leads to a negative result, we set the value to 0 before calculating the CDF.}  rapidly decreases at low values of $\mu_{S}$, showing that while this estimator is unbiased that lack of bias comes at a serious cost with regards to the likelihood of your estimator being accurate given the observations. The mean value has an average CDF value that is slightly higher than $0.5$, with an increasing value as $\mu_{S}$ decreases.

In summary, our new probabilistic calculation has many ``desirable'' characteristics - it is positive definite, converges to zero sky counts as the expected background greatly exceeds the total number of observed counts, and converges to simple subtraction in the high counts limit. It has a better tendency to be representative of the underlying posterior PDF than simple subtraction.  We therefore conclude that the $\mu_{S}^{\star}$ estimator is superior to simple subtraction at low counts without introducing any adverse systematic effects at higher count rates.

\section{Creating Images}
\label{S::imagemake}

Our new calculations offer the ability to assign, for each event in a typical x-ray or gamma ray event list file, the probability of that event being associated with either a sky photon or the NXB. This enables background events to be removed or down-weighted from images even in complex situations where the telescope is in continuous motion during the observation. In this section we discuss how to create meaningful images using these calculations for both pointed observations and survey data.  There are several different variants of a standard image that are each well suited for different tasks and limited for others. 

\subsection{Random Removal of Likely Background Events} 
\label{S::statbgsub}
The background events from an event list can be removed statistically by assigning, to each event, its associated value of $p_{S}$. A second number between 0 and 1 is randomly drawn from a uniform distribution. If this random number is less than the value of $p_{S}$ then the event is kept and included in the final image. Otherwise it is discarded. For a survey telescope like ART-XC, the surviving events are mapped to sky coordinates using the spacecraft's attitude at the event's arrival time and divided by the vignetting factor of the telescope at the event's detector coordinates. The total image of vignetting corrected events can then be divided by an exposure map that describes the dwell time of the telescope at each position to create a fully exposure corrected image. This method of background subtraction offers an exposure-corrected image that is ``flat'', in the sense that the large-scale spatial distribution of counts should be corrected for the expected behavior of the optics/detectors and pointing direction. It is limited in that it is susceptible to hot pixels that survive this removal process. For similar reasons, the resultant image may also have reduced signal-to-noise for strong sources, as events originating from ``real'' sources can possibly be removed from the image.

\subsection{A Fractional Photon Map} 
\label{S::fracphoton}
A different variant on the background subtracted image is to assign the probability $p_{S}$ as calculated above to each event's sky position. In essence, we assign to each event a fraction of a photon, the precise value of which is given by the probability of the event originating from a sky photon. These fractional photons can then be corrected for vignetting before being mapped to their sky position. This version has far more pixels with non-zero values than the statistically subtracted version described above - no events are removed outright. Because the background events are still nominally present, the spatial distribution of image pixel values will not trace the vignetting function of the optics.  This is one method of producing an image that has fewer noisy pixels and better preserves the flux of candidate sources, however. It may be well-suited for calculating the significance of any source flux at positions of interest. 

\subsection{A False Probability Map}
\label{S::falseprob}
For some applications, an image of the $p_{N}$ (i.e. the probability of a particular event being associated with an background event) values for each sky pixel may be of interest. This can be an especially useful quantity for source detection, as it can be easily be turned into a false probability map. For missions with multiple telescopes of similar performance such as ART-XC, the false probability value across all telescopes, $p_{F}$, can be determined as a simple product

\begin{equation}
    p_{F} = \prod_{j} p_{N, j} 
\end{equation}

where $p_{N,k}$ is the value of $p_{N}$ for the $j^{th}$ instrument or telescope\footnote{For pixels with zero events, this probability value should be unity}. Bright sources would appear as minima in the false probability map, so for visualization/source detection purposes it may be more appropriate to present an image of $1 - p_{F}$. Such a map can readily identify regions with anomalously high sky photons, but given that the image pixel values are not proportional to the event counts it cannot be corrected for the exposure map.

\section{Examples Using ART-XC Data}

The data from the recently launched ART-XC telescope are one example where such a probabilistic method of background subtraction is essential for producing meaningful images. Before discussing the nature of the data, we will summarize a few hardware and operational details for the ART-XC mission that motivate our analysis procedure. More details regarding the ART-XC detector hardware can be found in \cite{Pavlinsky2021}. 

\subsection{Telescope Summary}
The ART-XC telescope system is composed of seven separate mirror module assemblies (MMAs) and detector systems, known as the Unit of R\"{o}ntgen Detectors (URDs). Each URD is a CdTe detector with two perpendicular arrays of 48 Au/Pt electrode strips, resulting in a total of $48 \times 48 = 2,304$ detector ``pixels'' \citep{Pavlinsky2018_PartI} with an angular size of $\sim 45^{\prime \prime}$. Laboratory tests of the MMA's with high resolution CCD's have measured the on-axis half-power diameter (HPD) for each MMA to be $\sim 30^{\prime \prime} -40^{\prime \prime} $ \citep{Pavlinsky2021}, a value that is further degraded when convolved with the larger pixel size of the flight URD's. 

The integrated background count rate in the $4-12 \keV$ band for these detectors is estimated as $\sim 1.6 \times 10^{-4} \thinspace \mathrm{cts} \s^{-1} \thinspace \mathrm{pix}^{-1} $. This average value across all detector pixels was determined using pointed observations of a blank region of the sky \citep{Pavlinsky2021Cat}. Pixel and telescope-dependent structure in the background is clearly observed, with individual pixel values varying by as much as a factor of $\sim 2$ around this nominal value. We therefore expect $\sim 0.2 \thinspace \mathrm{cts} \s^{-1}$ across all of the active detector pixels for each telescope.

The data we are considering in this work is from ART-XC's all-sky survey, which covers the entire sky every six months. Its motion during the survey corresponds to 1.5 degrees per minute, or 90 arcseconds (two detector pixels) per second. From any particular source on the sky we should anticipate the ART-XC data only having a few counts during a pass dispersed over many detector pixels. A sizable fraction of the total counts in these detector pixels will originate from the background. These values should make it clear that a simple subtraction of the background in either sky or detector coordinates will be severely limited in this regime and is best accounted for on an event-by-event basis.

\subsection{Tests Using Simulated ART-XC Data}

We will first test the quality of our background subtraction using simulated ART-XC event lists where the positions and fluxes of sources as well as the input background spectrum are known \textit{a priori}. 
The ART-XC data is simulated using {\sc SIXTE} \citep{Dauser2019SIXTE}, which uses the ART-XC telescope specifications described in \citet{Pavlinsky2018_Overview}. The simulation also makes use of the latest ART-XC calibration files, including the vignetting functions, PSF, and the auxiliary response function\footnote{The full ART-XC calibration data base is in preparation by the ART-XC team, but some of the ground-based calibration performance can be found in \citet{Pavlinsky2018_PartI} and \citet{Pavlinsky2019_PartII}.}. 
The simulation makes use of the real ART-XC attitude pointings near the $\approx 200$ deg$^2$ North Ecliptic Pole (NEP hereafter) region from the first 6-month epoch of the ART-XC all sky survey, modified only to remove the gaps in coverage times inherent to the real ART-XC observations. Images for all of these simulations are created by calculating a gnomonic projection of the celestial coordinates (right ascension $\alpha$ and declination $\delta$) centered at the J2000 position $\alpha, \delta = 266.53846^{\circ}, 66^{\circ}$ with an image pixel size of $15^{\prime\prime}$.  Simulations were only run for one of the seven telescopes.

We choose to simulate an extended source whose morphology is based on the \textit{Chandra} image of the galaxy cluster Abell 2146. The purpose of this set of simulations is to examine the performance of the background-subtraction algorithm instead of mimicking a specific X-ray source. We assume the cluster spectrum is a simple isothermal model with Galactic absorption ( {\sc wabs}$\times${\sc vapec} in {\sc XSPEC}). The Galactic absorption column density is fixed to $1\times10^{21}$cm$^{-2}$, and the cluster temperature is fixed at  $kT = 6 \keV$. The metallicity is fixed to $Z = 0.3 \thinspace Z_{\bigodot}$ and the redshift is fixed to $z=0.1$. The normalization of this thermal model was adjusted such that our simulated cluster has a $4-30 \keV$ flux of $1 \times 10^{-11} \ergpcmsqps $.

\begin{figure} 
    \centering
    \includegraphics[width=0.45\textwidth]{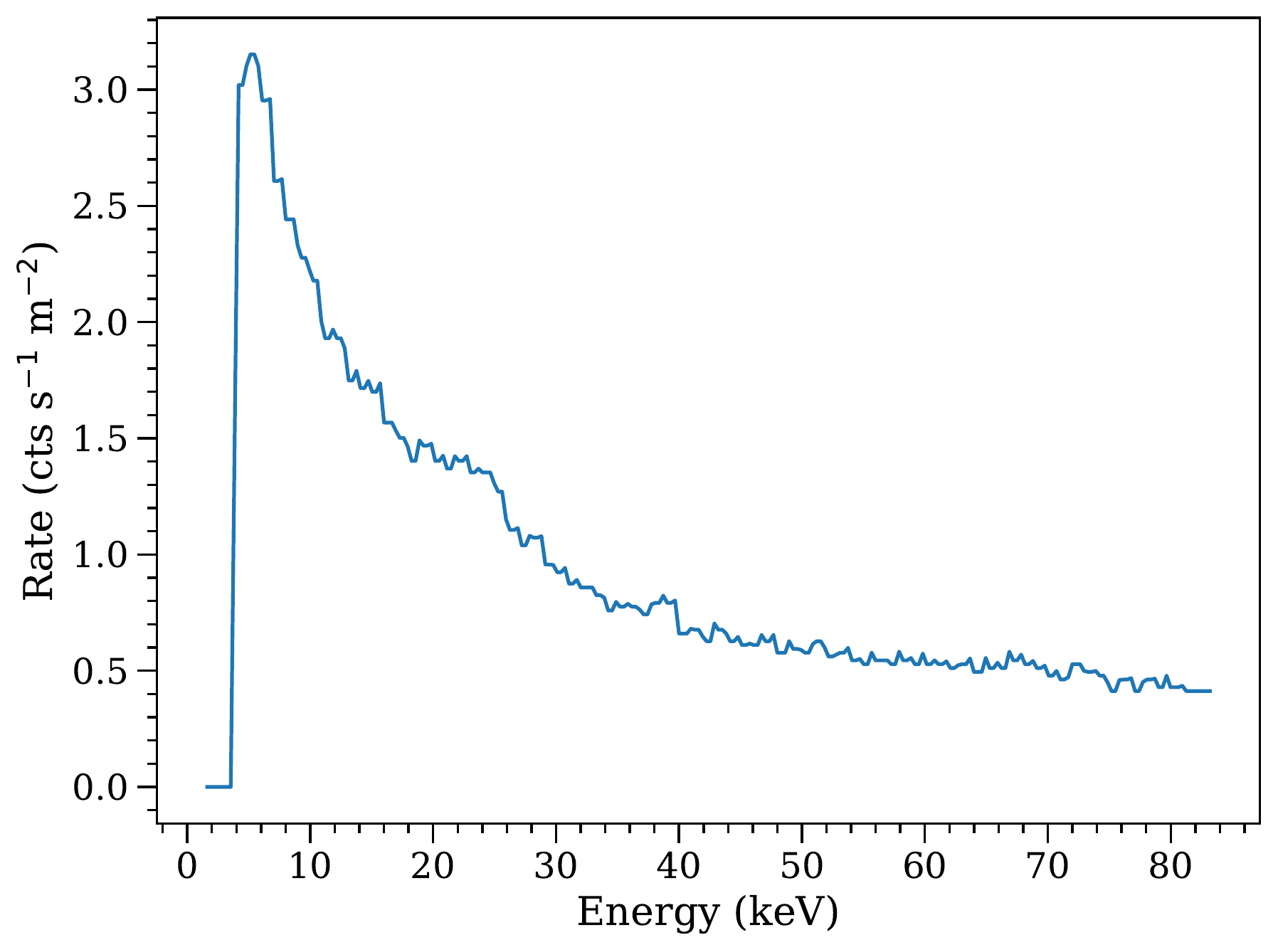}
   \caption{The background counts spectrum used as input for simulated ART-XC images. 
   }
    \label{fig::ARTXC_SimSpec}
\end{figure}

We have run simulations of this cluster both with and without an injected background spectrum. The background spectrum, shown in Figure \ref{fig::ARTXC_SimSpec}, was derived from $\approx 622 $ ks real ART-XC observations of sky regions without known X-ray sources.  A third simulation that has this same input background spectrum but no cluster emission was also run. A total of 358 and 159 events were detected within a $3^{\prime}$ radius of the cluster center for the cluster+background and background simulations, respectively. From these values, we should expect any source flux that accurately accounts for the background to contain approximately $\sim 200$ net counts.  A larger $27^{\prime}$ radius region located far away from the cluster had a total of 831 and 794 counts in the cluster+background and background simulations, suggesting the two event lists have background count rates similar at the $\sim 5\%$ level. 

For the cluster+background simulation, we apply our background subtraction algorithm to the output event list in $5 \times 5$ square regions of sky pixels ($1.25^{\prime} \times 1.25^{\prime}$). Because of the continuous motion of the telescope and frequent passes through the NEP during the survey ($\sim 6$ times per day), the exposure map (and subsequently the background counts map) is extremely non-uniform in sky coordinates. Our expectation value for the background counts in each of these $5 \times 5$ pixel regions is determined using the identical region of the sky in the background-only simulation. A single value of $\mu_{S}^{\star}$ was determined for each of these regions, which is used to assign an equal value of $p_{S}$ to every event contained therein. We used those values of $p_{S}$ to construct new images of the cluster - one using the random removal of background method and the other projecting fractional photons back onto the sky.

\begin{figure*} 
    \centering
    \includegraphics[width=0.85\textwidth]{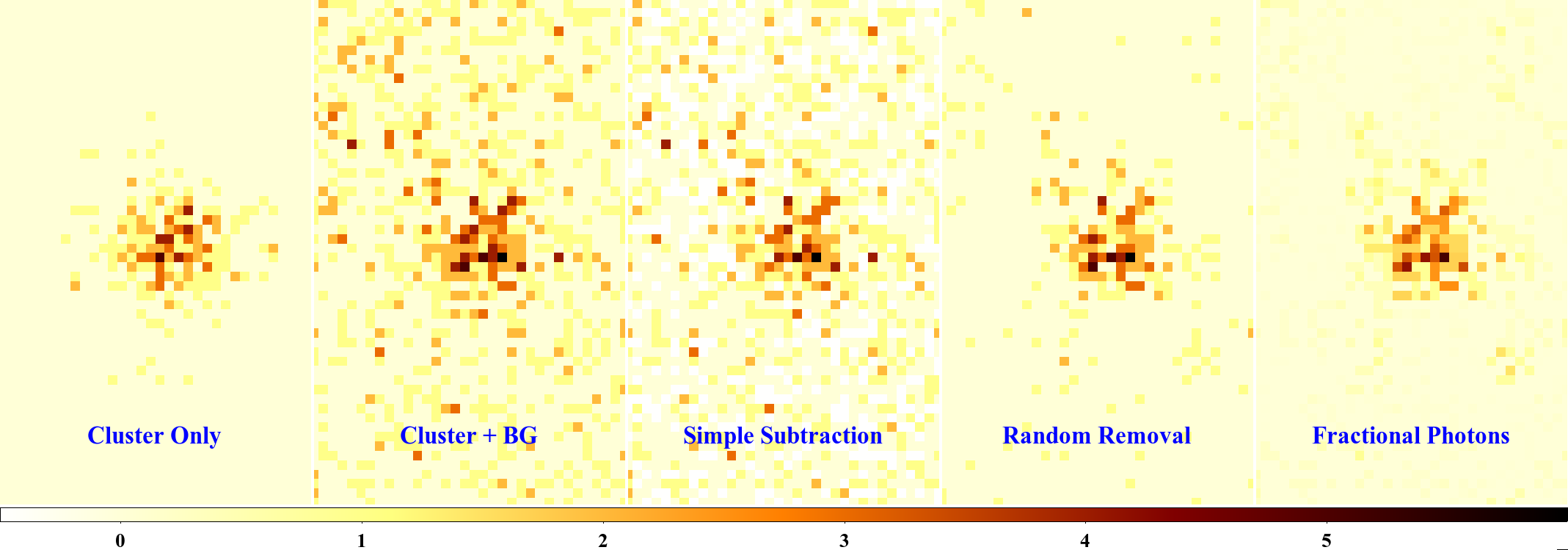}
   \caption{Simulated ART-XC images of a galaxy cluster with and without background subtraction. The color bars and scales for all five images are identical, with white pixels corresponding to negative counts.   
   Far left: The image of the galaxy cluster derived from a simulation without any input background. 
   Middle-left: The image made from the simulated cluster + background event list. No background subtraction was applied to this image.
   Middle: The ``clean'' image, made from subtracting the background image derived from a simulation without cluster emission from the image derived from the cluster + background simulation.
   Middle-right:    The image made from the simulated cluster + background event list with likely background events removed randomly from the image. 
   Far-right: The image made from the simulated cluster + background event list with all events down-weighted by the value of $p_{S}$. 
   All three images with background treatment used identical background data. }
    \label{fig::ARTXCimg}
\end{figure*}

Images of the cluster with and without our probabilistic subtraction calculation applied are shown in Figure \ref{fig::ARTXCimg}. For comparative purposes, we have also produced an image of the cluster with the background image subtracted simply.  Using simple aperture photometry within this $3^{\prime}$ radius, the random removal and fractional photon images have a signal-to-noise ratio of $\sim 10.78$ and $11.1$, respectively, as opposed to $\sim 8.8$ without any background subtraction. For simple subtraction, the nominal signal-to-noise ratio is $\sim 14.9$, but this value is artificially inflated by the presence of unphysical negative pixels biasing the residual background low. A simple clipping to set all of the negative image pixel values to $0$ results in a signal-to-noise ratio nearly identical to the case without any background subtraction. 

\begin{figure*} 
    \centering
    \includegraphics[width=0.75\textwidth]{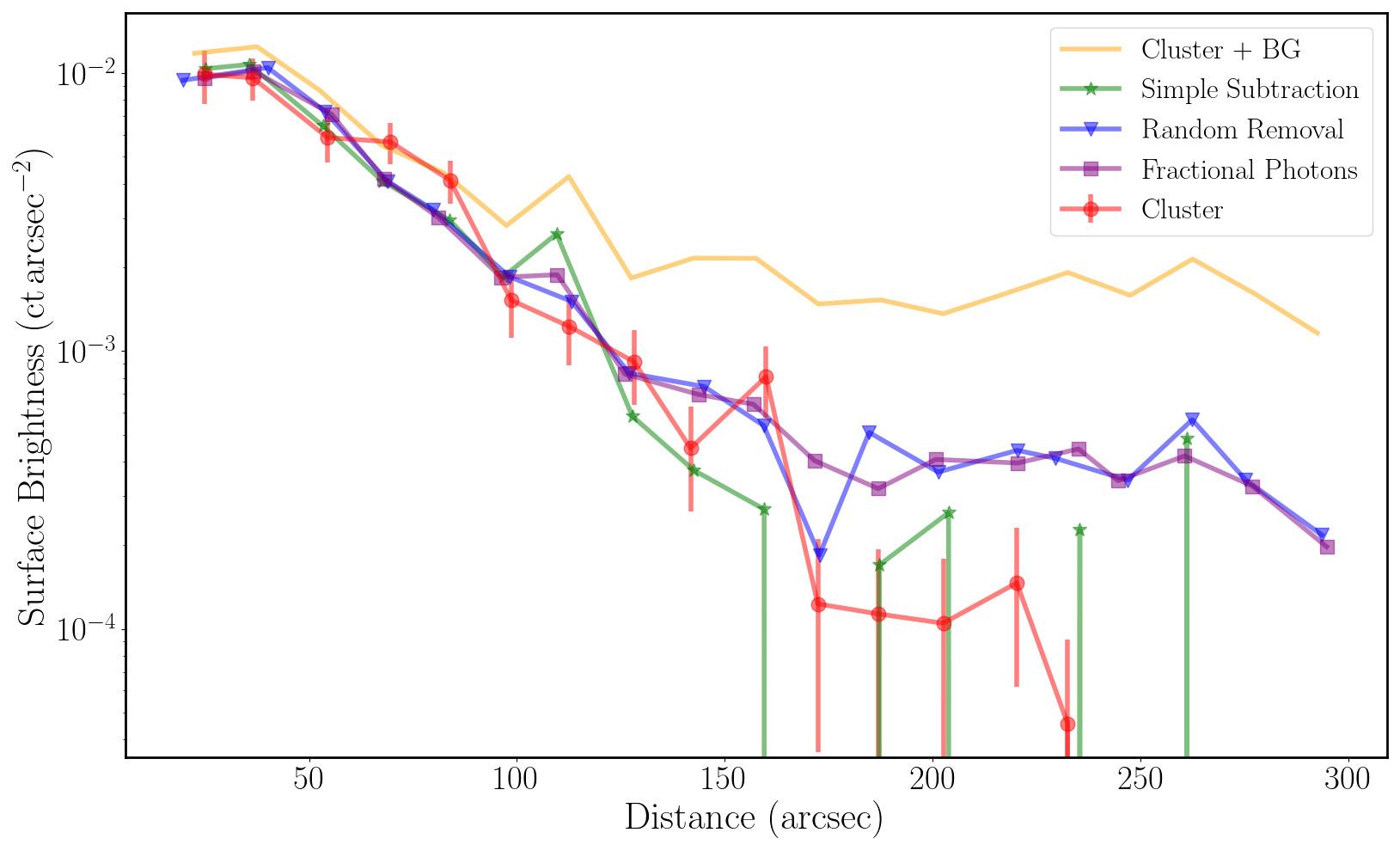}
   \caption{The surface brightness profiles for each of the five images shown in Figure \ref{fig::ARTXCimg}.  To guide the reader as to the expected statistical fluctuations in the curves presented here, simple Poisson error bars ($\sqrt{N}$) have been included on the cluster only surface brightness profile. The markers for each of these curves have been randomly offset from each other by up to $\pm 3^{\prime \prime}$ exclusively for presentation purposes. Both the random removal and fractional photon methods enable the cluster emission to be resolved out to larger distances than images with or without simple background subtraction. 
   }
    \label{fig::A2146_SB}
\end{figure*}

As a final test of the robustness and accuracy of our probabilistic subtraction algorithm, we show the resultant surface brightness profiles for the five images in Figure \ref{fig::A2146_SB}. From these surface brightness profiles it is clear that both the random removal and fractional photon images do not greatly affect the extended structure of the cluster on average and gives a positive definite value at all radii, unlike the simple subtraction. Furthermore, the absence of any over-subtraction in our new methods enables the surface brightness of the cluster to be resolved at larger radii than what is possible with the simple subtraction curve.  Comparing the random removal and fractional photon surface brightness profiles to that without any background subtraction at distances of $\gtrsim 5 ^{\prime}$ , we find that the average ``background'' surface brightness is reduced by factor of $\sim 5$.

\subsection{Tests Using Real ART-XC Data}

After demonstrating the viability of this background subtraction calculation with simulated data of a known source and background spectrum, we repeat the test using real observations using the ART-XC data. The input event lists, one for each of the seven telescopes, are six months of integrated exposure throughout a $ 6.92^{\circ} \times 3^{\circ}$ rectangle centered  at a position ($\alpha, \delta = 273.46154^{\circ},  66.0^{\circ}$) near the NEP. Final images are produced by summing up the counts across all seven telescopes and dividing by the summed vignetting-corrected exposure map. The estimated background is determined by using the results of the ``blank-sky'' observations taken by ART-XC \citep{Pavlinsky2021Cat}, renormalized to match the $30-70 \keV$ count rate observed in each of the constituent daily event list files that, when combined, produce the integrated event list.   The mirrors have zero effective area at these energies, meaning that all events observed with these energies are associated with the background. Similar to what was done for the simulated data above, we project the expected detector background counts onto its corresponding sky position using the instantaneous pointing of the telescope at every second during this six month exposure. Our probabilistic background subtraction is subsequently performed using the background maps and observed data in $1.25^{\prime}$ square regions on the sky. 

\begin{figure*} 
    \centering
    \includegraphics[width=0.75\textwidth]{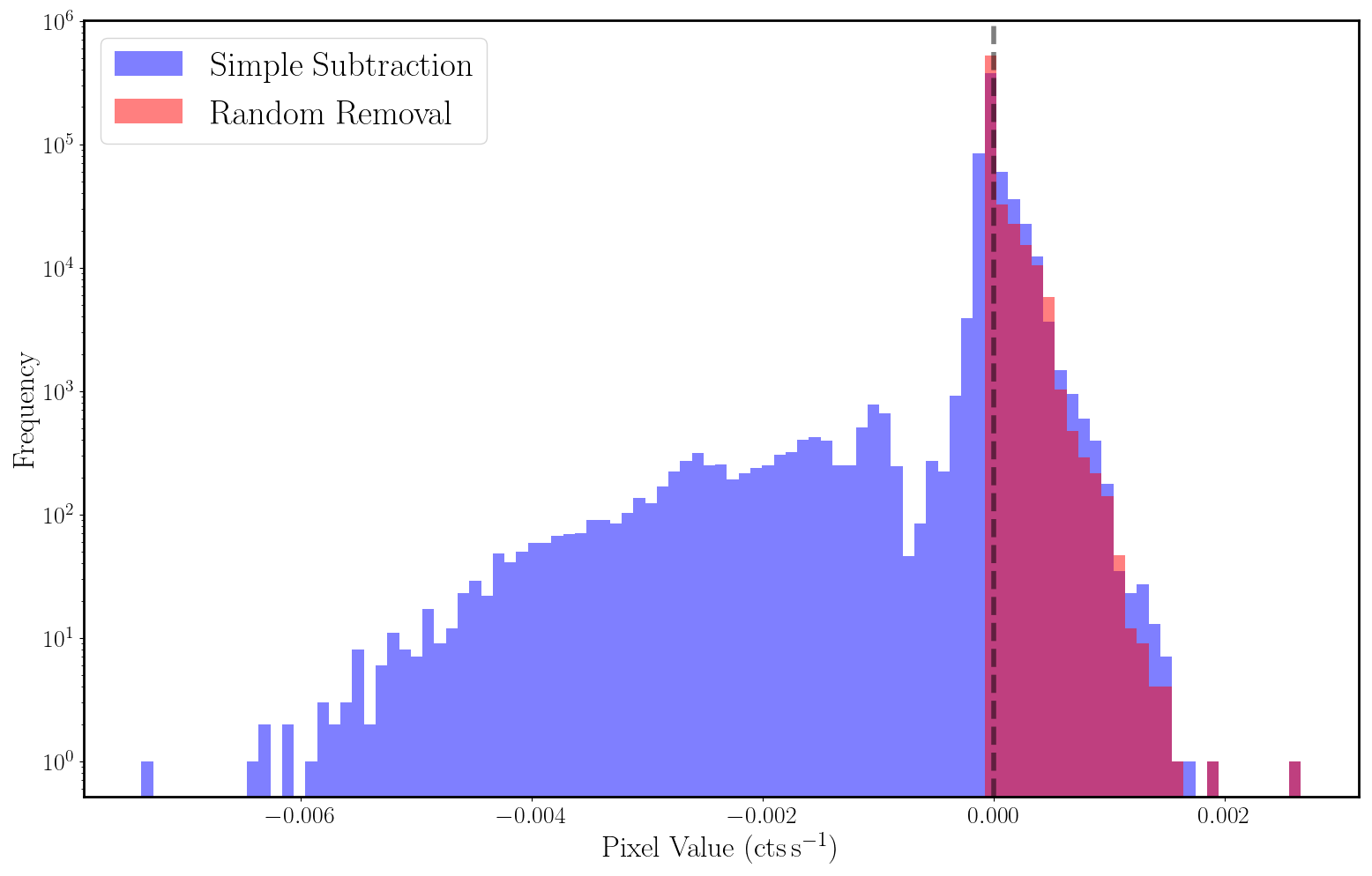}
      \includegraphics[width=0.95\textwidth]{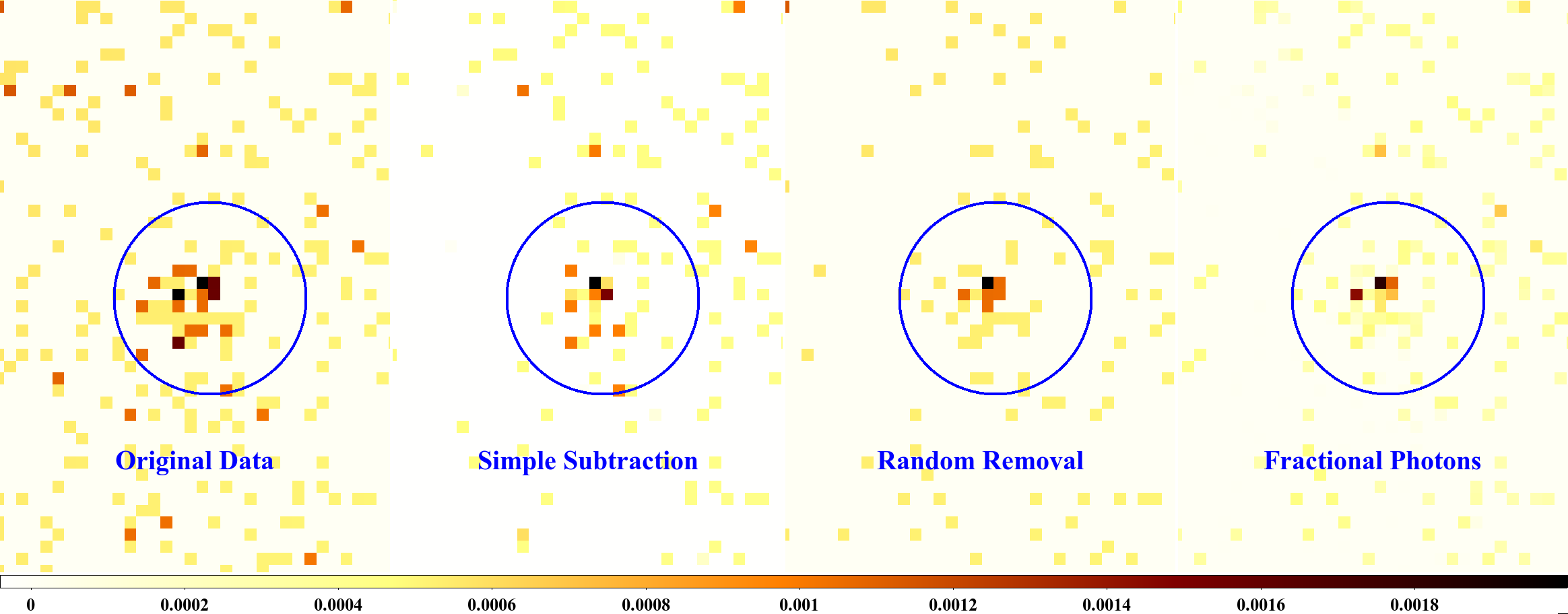}
   \caption{ Imaging results derived from ART-XC observations of a $6.92^{\circ} \times 3^{\circ} $ tile region near the North Ecliptic Pole. The images from each of the seven telescopes are individually corrected for background and the divided by the sum of the seven corrected exposure maps.  \textit{Top: } The distributions of pixel values across the entire tile using simple background subtraction (in blue) and the random removal method described in this work (in red). The dashed vertical line denotes zero counts.   \textit{Bottom: } Zoomed in images around one of the sources in the ART-XC first year source catalog, the star HD 170527. The scaling and color bars are identical for all four images, and the white color denotes pixels with negative values. The blue circle denotes a $2^{\prime}$ circle around the optical position of this star. From left to right, the four sub-panels show: the image of the source plus background; the simply subtracted image using the re-scaled blank sky data projected onto the sky using the spacecraft attitude history for this region; our new method of random removal of likely background events; and the image created by down-weighting each event by its value of $p_{S}$. The random removal and fractional photon methods use the exact same background sky map as the simple subtraction method. 
   }
    \label{fig::Tile273024_RealBGSub}
\end{figure*}

We demonstrate the effectiveness of this algorithm in two ways in Figure \ref{fig::Tile273024_RealBGSub}. The first is to show the distribution of pixel values across the full tile after applying various methods of subtracting background. The second demonstrates a zoomed-in image surrounding a real source detected in the ART-XC first year survey catalog: the star HD 170527. We perform a aperture photometry exercise similar to that performed for the simulated cluster emission, after renormalizing the exposure corrected image (units of $\mathrm{cts} \s^{-1}$) by multiplying the data near the star by the average value of the exposure map (units of $\s $) in that same region. With no background subtracted/removed the signal-to-noise ratio for this source is $5.15$. The signal-to-noise ratios for simple subtraction, random removal, and fractional photons are  $5.06$, $5.51$, and $5.06$ respectively. The gains in signal-to-noise for this case are not as significant as for an extended source, but there is nevertheless no evidence that our new methods of suppressing background are adversely affecting the image statistics. For fainter sources at lower signal-to-noise even the modest improvements in significance can greatly improve the robustness and reliability of any candidate sources. 

\section{Discussion}
The method of accounting for background events presented here, while originally developed and designed to produce exposure-corrected images out of the ART-XC survey data, can be applied to data from any ``photon counting'' observatory. The particulars of the implementation for each mission will only vary in how the background event rate is estimated in a detector region, what that rate is, and how the telescope converts between detector and sky coordinate systems. 

The treatment of background events described in this work has several properties that compare favorably to a simpler background subtraction. First and foremost, the results of the estimated sky photon count rates are ensured to be positive definite even when the expected number of background events in a pixel exceeds the total number of observed counts. The positive definite nature of the resultant image is both more physically sensible and aesthetically pleasing when an output image is created. In particular, no artificial cuts or restrictions on the image data are needed to address negative values in the image pixels. Importantly, tests with simulated data show that this weighted average provides an unbiased estimate of the true net counts and converges to the intuitive value of zero when the expected background counts become much larger than the total number of counts observed. 

The mathematical framework described here may also have applications with other sources of ``background'' beyond the non-photon background for which this algorithm was specifically developed. While we have discussed this algorithm primarily in terms of separating background events originating from charged particles and electronic noise from sky photons in ART-XC data, any background in an event list that can be estimated independently can be accounted for using these same calculations, at least in principle. As one example, the variable $\mu_{N}$ could instead correspond to the quiescent flux of a flaring source, with $\mu_{S}$ corresponding to the flux of flares over a specific time period. In such a case, the flux of the flare can be robustly estimated while accounting for the expected Poisson fluctuations in the quiescent count rate in each individual time bin. We caution that, for any application of this algorithm, the interpretation must account for the inherently asymmetric and non-Gaussian distributions of derived parameter values. 

One caveat for this method is that it does not explicitly account for uncertainties or outright errors in the expected background counts per pixel, $\mu_{N}$. Because these calculations are designed to operate in the regime where the total counts per pixel are subject to large statistical fluctuations, it is certainly plausible that the uncertainties in the expected background counts per pixel may be small relative to the other uncertainties in the calculation. We nevertheless encourage any users who create images using these formulae to test the results using a realistic distribution of background count estimates in order to quantify their impact on their final images. 

While the resultant images may be useful for identifying candidate sources that may not be apparent without background subtraction, this algorithm is not explicitly designed for such a purpose. The calculation presented here does not utilize any crucial telescope-specific information such as the point spread function (PSF) that is essential to properly quantifying the significance of the detection. In that sense, this algorithm is better suited for imaging extended, diffuse sources much larger than then PSF. Our tests using ART-XC data of a simulated galaxy cluster and a real star confirm that the signal-to-noise of the cluster emission was more significantly boosted by our improved methods of accounting for the background.  Any candidate source detected in any image (including the new calculations discussed in this work) must be subject to rigorous follow-up analysis that calculates their overall significance above the expected background counts within a region. We emphasize that most rigorous calculations of source significance, including those utilizing ART-XC data \citep[e.g.][]{Pavlinsky2021Cat} do not remove the background events at all but rather model it simultaneously with any candidate source counts. 

We caution that the practical details of applying this calculation to an X-ray or gamma ray telescope's data will depend strongly on the details of the telescope, even though the underlying statistical principles do not. Any user hoping to take advantage of this new calculation will need to make sure they are properly estimating and modeling the background structure (which may include its dependence on time, energy, detector coordinates, etc.. ) to sufficient accuracy. The choice of estimating the net counts in discrete regions of the sky for our ART-XC examples is largely driven by the presence of rich structure in the exposure map arising from the continuous motion of the spacecraft throughout the survey tile and the fact that each pass through the tile may have a different NXB count rate. This method of applying our new background treatment algorithms may apply to other telescopes "out of the box", but other means of binning the data may be required. Whatever method is ultimately developed for a specific mission, it must account for all of the unique aspects of their respective detector hardware and operations.  

Recent work \citep{Wilkins2020,Grant2020} using simulations of the upcoming ATHENA mission has shown that machine learning methods may be able to independently estimate the probability that an individual event is associated with the background. The methods presented here can be considered complimentary to those, in the sense that the our framework can be used in situations where such high fidelity simulations of X-ray photon interactions with telescope hardware are not available or feasible. This statistical method could also be used to help support or refute the event-specific probabilities determined by the machine learning algorithms. The methods for producing images discussed in \S \ref{S::imagemake} derived from the event-specific probabilities can be readily utilized in the same manner regardless as to how that probability was determined.  

\section*{Acknowledgements}
This work was supported by the National Aeronautics and Space Administration’s (NASA’s) Astrophysics Division, Science Mission Directorate.   The Mikhail Pavlinsky ART-XC telescope is the hard X- ray instrument on board the \textit{SRG} observatory, a flagship
astrophysical project of the Russian Federal Space Program
realized by the Russian Space Agency in the interests of the
Russian Academy of Sciences. The ART-XC team thanks
the Russian Space Agency, Russian Academy of Sciences,
and State Corporation Rosatom for the support of the \textit{SRG}
project and ART-XC telescope. The majority of the integrals and gamma function calculations we utilize in this work were investigated and verified using Wolfram Alpha \citep{WolframAlpha}. We finally thank the referee, Phillipe Laurent, for providing very helpful and constructive comments.

\section*{Data Availability}
The statistical calculations and simulated cluster data underlying this article will be shared on reasonable request to the corresponding author. The ART-XC survey data are currently not available to public, but the development of a publicly available archive of ART-XC data is currently ongoing. 

\bibliographystyle{mnras} 
\input{reference_defs}
\bibliography{NXBProbability}

\appendix 

\section{Convergence of $\mu_{S}^{\star}$ to Simple Subtraction in High Signal-to-Noise Limit}
\label{app::converge}
We will now demonstrate that our new estimator $\mu_{S}^{\star}$ is identical to a simple subtraction calculation ($k -\mu_{N}$) in the limit where $k \gg \mu_{N}$. The summary statistic described in Equation \ref{eq::mustar} 
\begin{equation}
\mu_{S}^{\star} = \frac{1}{V_{1}}\sum_{k_{N} = 0}^{k} \frac{(k - k_{N}) e^{-\mu_{N}} \mu_{N}^{k_{N}}}{k_{N!}} 
\end{equation}

where the normalization $V_{1}$ is defined as

\begin{equation}
    V_{1} = \sum_{k_{N} = 0}^{k} \frac{ e^{-\mu_{N}} \mu_{N}^{k_{N}}}{k_{N!}} 
\end{equation}

can be expressed exactly and analytically as

\begin{equation}
\mu_{S}^{\star} = \frac{e^{-\mu_{N}} \left[ \mu_{N}^{k+2} - e^{\mu_{N}}\left( -k + \mu_{N} -1 \right) \Gamma_{U}(k+2,\mu_{N}) \right]}{\Gamma(k+2)} - \frac{\Gamma_{U}(k+1, \mu_{N})}{\Gamma(k+1)}. 
\end{equation}

In this equation $\Gamma(x) $ is the gamma function, the real-valued extension of the factorial function with  $\Gamma(k+1) = k!$ when $k$ is an integer. When $k \gg \mu_{N}$ we can approximate the incomplete upper gamma functions with their respective gamma functions (in other words $\Gamma_{U}(k,\mu_{N}) \approx \Gamma(k)$ when $k \gg \mu_{N}$). Distributing the leading exponential factor to the two terms in the square brackets, canceling out the gamma functions in the numerators and denominators in this approximation limit, and simplifying leads to  

\begin{equation}
    \mu_{S}^{\star} \approx \frac{e^{-\mu_{N}} \mu_{N}^{k+2}}{\Gamma(k+2)} +k - \mu_{N}. 
\end{equation}

In this limit ($k \gg \mu_{N}$), this first term also converges to zero quickly enough to be negligible since $\Gamma(k+2) = (k+1)!$. In this regime the only terms that therefore survive lead to desired result of $\mu_{S}^{\star} \approx k - \mu_{N}$. 

\section{Bias of Simple Subtraction}
\label{app::bias}
We wish to show that simple subtraction as discussed in this paper ($k-\mu_{N}$) is an unbiased estimator of $\mu_{S}$. Using the operation $E(x) $ to denote the expectation value of a variable $x$, this is equivalent to showing that $E(k - \mu_{N}) = \mu_{S}$. We note that since $k$ is a random variable that follows a Poisson distribution from the sum of our sky and background terms, its expectation value is given as 

\begin{equation*}
    E( k ) = \mu_{S} + \mu_{N}  
\end{equation*}
\noindent Since the expectation value operation is linear and the expectation value of a distribution parameter is the parameter itself, we can then rewrite our expectation value as 
\begin{equation*}
    E(k - \mu_{N} ) = E(k) - E(\mu_{N}) = \mu_{S} + \mu_{N} - \mu_{N} = \mu_{S}
\end{equation*}
\noindent Thereby proving that the simple subtraction calculation is unbiased.

\bsp	
\label{lastpage}
\end{document}

%% file: units.tex
\def\cm{{\rm\thinspace cm}}
\def\erg{{\rm\thinspace erg}}

\def\keV{{\rm\thinspace keV}}

\def\s{{\rm\thinspace s}}

\def\ergpcmsqps{\hbox{$\erg\cm^{-2}\s^{-1}\,$}}

%% file: reference_defs.tex
\def \aap {A\&A} 
\def \statisci {Statis. Sci.}
\def \physrep {Phys. Rep.}
\def \pre {Phys.\ Rev.\ E}
\def \sjos {Scand. J. Statis.} 
\def \jrssb {J. Roy. Statist. Soc. B} 


%

\def \araa {ARA\&A}
\def \aj {AJ}
 \def \aas {A\&AS}
\def \apj {ApJ}
\def \apjl {ApJL}
\def \apjs {ApJS}
 \def \aaps {Ap\&SS}
\def \mnras {MNRAS}
\def \nat {Nat}
 \def \pasp {PASP}
\def \sovast{Soviet~Ast.}
\def \gca {Geochim.\ Cosmochim.\ Acta}
 \def \icarus {Icarus}
\def \prd {Phys.\ Rev.\ D}
\def \prl {Phys.\ Rev.\ Lett.}
\def \pss {Planetary \& Space Science }
\def \planss {Planetary \& Space Science}
 \def \posp {Proceedings of SPIE}
\def \procspie {Proceedings of the SPIE}